\documentclass[iop]{emulateapj}
\usepackage{graphicx}

\usepackage{color}

\usepackage{epstopdf}
\usepackage{amsmath}

\def\ltsima{$\; \buildrel < \over \sim \;$}
\def\simlt{\lower.5ex\hbox{\ltsima}}
\def\gtsima{$\; \buildrel > \over \sim \;$}
\def\simgt{\lower.5ex\hbox{\gtsima}}

\newcommand\lsim{\mathrel{\rlap{\lower4pt\hbox{\hskip1pt$\sim$}}
\raise1pt\hbox{$<$}}}
\newcommand\gsim{\mathrel{\rlap{\lower4pt\hbox{\hskip1pt$\sim$}}
\raise1pt\hbox{$>$}}}

\usepackage{color}

\shorttitle{DM signatures of SMBH binaries }
\shortauthors{Naoz et al.}

\begin{document}

\title{Dark Matter Signatures of Supermassive Black Hole Binaries}

\author{Smadar Naoz$^{1,2}$, Joseph Silk$^{3,4,5}$ and Jeremy D. Schnittman$^{6,7}$}
\affiliation{
 $^1$ Department of Physics and Astronomy, University of California, Los Angeles, CA 90095, USA. \\   $^2$ Mani L. Bhaumik Institute for Theoretical Physics, UCLA, Los Angeles, CA 90095, USA \\ 
 $^3$ Institut d'Astrophysique de Paris, UMR 7095 CNRS, 
 Sorbonne Universit\'es,
98 bis, boulevard Arago, F-75014, Paris, France \\ $^4$  The Johns Hopkins University, Department of Physics and Astronomy, Baltimore, Maryland 21218, USA  \\ $^5$ Beecroft Institute of Particle Astrophysics and Cosmology, University of Oxford, Oxford OX1 3RH, UK \\
$^6$ NASA Goddard Space Flight Center, Greenbelt, MD, 20771 \\
$^7$ Maryland Joint Space-science Institute, College Park, MD 20742
   }%

\begin{abstract}
A natural consequence of the galaxy formation paradigm is the existence of supermassive black hole (SMBH) binaries. Gravitational perturbations from a  far-away SMBH companion can induce high orbital eccentricities on dark matter particles orbiting the primary SMBH via the eccentric Kozai-Lidov mechanism.  This process yields an influx of dark matter particles into the primary SMBH ergosphere, where test particles linger for long timescales. This influx results in high self-gravitating densities, forming a dark matter clump that is extremely close to the SMBH. In such a situation, the gravitational wave emission between the dark matter clump and the SMBH is potentially detectable by LISA. If dark matter self-annihilates,  the high densities of the clump will result in a  unique co-detection of gravitational wave emission and high energy electromagnetic signatures. 
\end{abstract}

\maketitle

\section{Introduction}
The hierarchical nature of the galaxy formation paradigm suggests that major galaxy mergers may result in the formation of supermassive black hole (SMBH) binaries \citep{DiMatteo+05,Hopkins+06,Robertson+06,Callegari+09}. While observations of SMBH binaries are challenging, there are several observed binary precursors with sub-parsec  to tens to hundreds of parsec separations 
 \citep[e.g.,][]{Sillanpaa+88,Rodriguez+06,Komossa+08,Bogdanovic+09,Boroson+09,Dotti+09,Deane+14,Liu+17,Bansal+17,Kharb+17,Runnoe+17,Pesce+18,Guo+19}.  
  Furthermore, several observations of active galactic nuclei pairs with kpc-scale  separations have been suggested as SMBH binary precursors  \citep[e.g.,][]{Komossa+03,Bianchi+08,Comerford+09bin,Liu+10kpc,Green+10,Smith+10,Comerford+18}.
   Numerical {\bf simulations} for spheroidal gas-poor galaxies suggest that these binaries can reach parsec separation and may stall there \citep[e.g.,][]{Begelman+80,Milosavljevic+01,Yu02}.

While the Dark Matter (DM) distribution in galaxies has been studied extensively in the literature, the DM profile for sub-kpc scales is largely  unknown. In \citet{NaozSilk14}, we suggested that gravitational perturbations in SMBH binaries can have important implications for the DM distribution around the {\it less} massive member of the binary. The requirement that the perturbing SMBH will be more massive than the primary arises from the need to overcome general relativistic precession of the DM particle orbits \citep{Naoz+12GR}. Gravitational perturbations from a far-away SMBH, on a DM particle orbiting around the primary SMBH can result in extremely high eccentricities due to a physical process known as the ``Eccentric Kozai-Lidov'' (EKL) mechanism \citep{Naoz16}. The eccentricities can reach extreme values \citep{Li+13} such that the pericenter passage of the DM particle reaches the SMBH ergosphere (or even the event horizon) \citep{NaozSilk14}. This process results in a DM torus-like configuration around the less massive SMBH \citep{NaozSilk14}. These torus particles were initially in a less favorable EKL regime of the parameter space, compared to those that reached high eccentricities.  

\begin{figure}
\centering
\includegraphics[width=\linewidth]{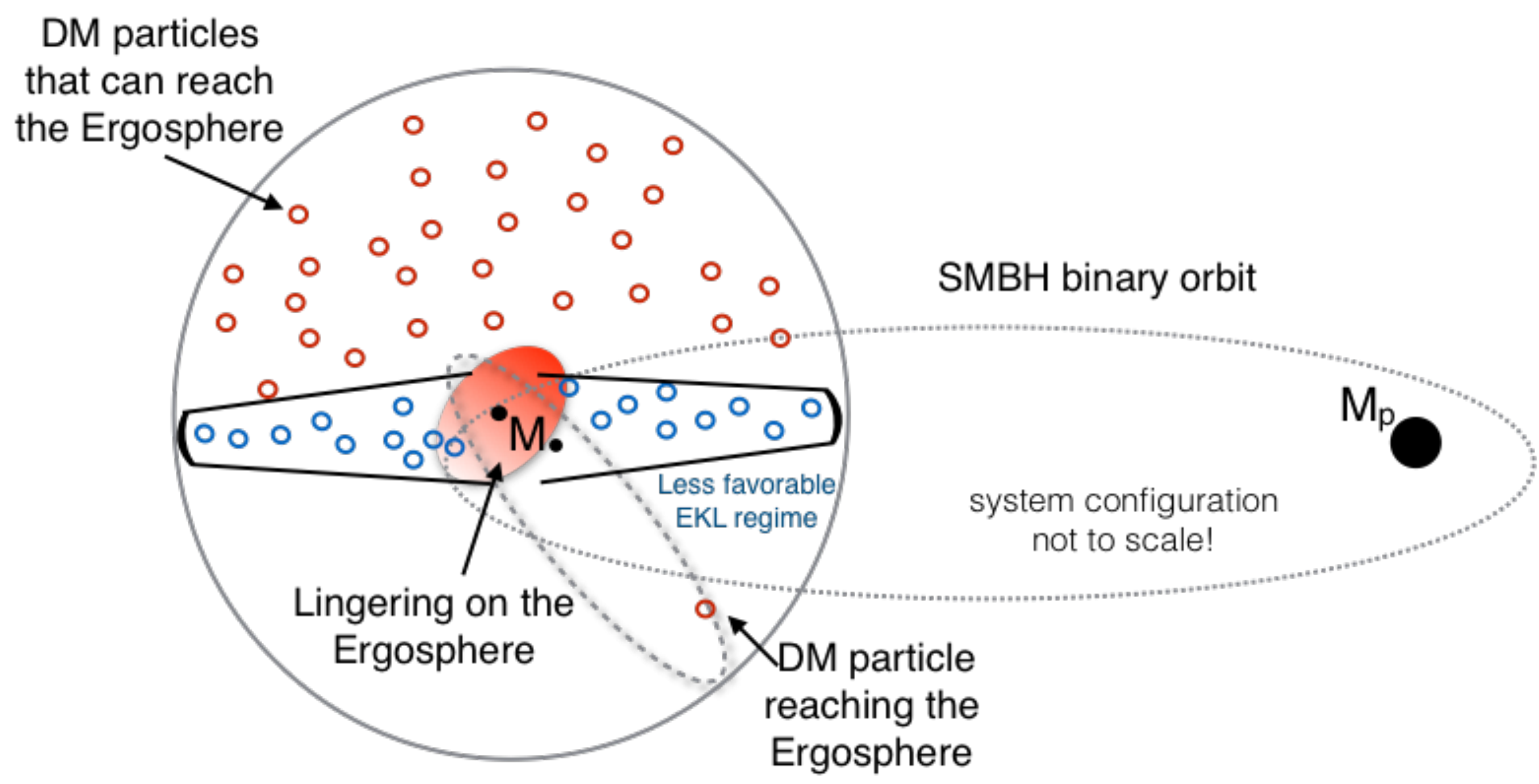}
\caption{ {\bf An illustration of the configuration}. The SMBH spin  can have an arbitrary orientation. 
  }\label{fig:cartoon}
\end{figure}

The low-energy, low angular momentum orbit of a test particle around a spinning black hole has been addressed in the literature up to   4th order in the post-Newtonian approximation \citep{Will+17}, and may yield an increase of the DM density around a rotating SMBH  \citep{Ferrer+17}.  The EKL mechanism in SMBH binary systems results  in  extremely low angular momentum orbits of the DM particles \citep{NaozSilk14}. These particles spend a significant part of their orbit zooming around the ergosphere, before continuing with their orbit \citep[e.g.,][and see below for more details]{Schnittman15}.

Here we show that even the temporary accumulation of DM in the ergosphere of an SMBH, as a result of zoom-whirl orbits, can reach such high densities as to allow for formation of self-gravitating DM clumps. Such a clump then emits gravitational waves (GW), potentially detectable by LISA, while possibly undergoing self-annihilation. This process may yield  a unique co-signal of GW emission and high energy electromagnetic signature arising from the self-annihilation process of DM (see Fig.~\ref{fig:cartoon}).

\section{Self-Gravitating DM Clumps}
DM is expected to be inhomogeneous and clumpy \citep[e.g.,][]{Silk+93,Berezinsky+10,Berezinsky+14}. This clumpy nature can be explained as a simple extrapolation to very small scales of the primordial power spectrum, and in large parts of the Universe these clumps are expected to be free from gas \citep[e.g.,][]{Naoz+14SIGOs,Popa+16,Chiou+18,Chiou+19}.  Furthermore, some DM clumps may have formed shortly after or during radiation-matter equality due to phase transitions, topological defects, or collapse into primordial perturbations \citep[e.g.,][]{Kolb+94,Starobinskij92,Berezinsky+10}. Moreover, these clumps may have formed at  earlier epochs due to accretion onto primordial black holes \citep[e.g.,][]{Bertschinger85,Ricotti09,Lacki+10,Eroshenko+16,AliHamoud+17}. Regardless of their formation mechanism, these clumps need to be self-gravitating in order to resist disruption from other objects in the Universe.

\begin{figure}
\centering
\includegraphics[width=1\linewidth]{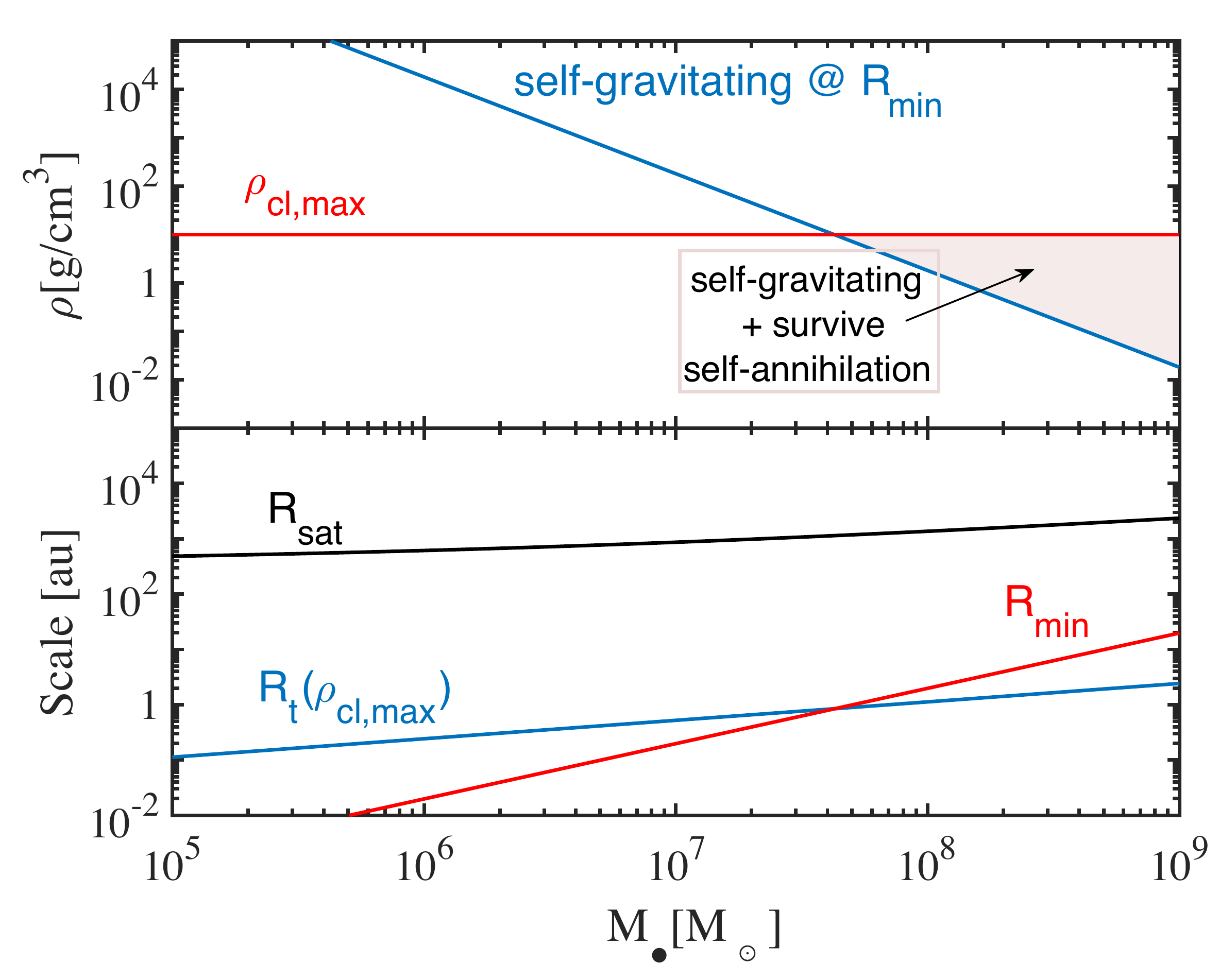}
\caption{ {\bf Typical density and physical scales in the system}. {\it Top panel:} we consider the self-gravitating density required at $R_{\rm min}$ (blue line).  We also show (red, horizontal line) the  maximum clump density for a self-gravitating, self-annihilating DM clump within a dynamical time-scale (adopting $m_\chi=100$~Gev DM particle). {\it Bottom panel:} relevant physical scales in the system. We consider the ergosphere scale, $R_{\rm min}$ (Eq.~(\ref{eq:Re})), as well as the tidal radius, $R_t$, (Eq.~\ref{eq:Rt}), for $\rho_{\rm cl}=\rho_{\rm cl,max}$. {We also show $R_{\rm sat}$, the saturated radius of the DM density, due to self-annihilation, see section \ref{sec:clump}. }
}\label{fig:scales}\vspace{0.1cm}
\end{figure}

In the vicinity of a SMBH with  mass $M_\bullet,$ we  define the tidal radius, at which the gravitational tidal field of the SMBH overcomes the DM clump self-gravity\footnote{{New work by \citet{Gafton+19} showed that the radius of marginal disruption ($R_t$) does not change much under Kerr geometry. }}:
\begin{equation}\label{eq:Rt} 
R_t\sim \left(\frac{3M_{\bullet}}{4\pi \rho_{\rm cl}}\right)^{1/3} \ ,
\end{equation}
where $\rho_{\rm cl}$ is the density of the DM clump. In Naoz \& Silk, we 
 showed that gravitational perturbations from a distant SMBH can cause high eccentricity excitations to the DM particle orbits reaching all the way to the ergosphere radius. The ergosphere represents a special location for a spinning SMBH. Here, test particles, such as DM, cannot stay stationary with respect to an outside observer \citep{Misner+73}, and tend to linger for long time-scales \citep{Schnittman15}. {\it We note that the analysis of this section is agnostic to any mechanism that produces an over-density at the ergosphere and simply describes the requirement for such a clump to exist.}
 
 The ergosphere radius is  (depicted in Figure \ref{fig:scales}): 
\begin{equation}\label{eq:Re} 
R_{\rm min}=\frac{2 G M_{\bullet}}{c^2} \ ,
\end{equation}
where $c$ is the speed of light and $G$ is the gravitational constant. 
We find the critical density for a self-gravitating clump to remain bound at the ergosphere by setting $R_{\rm min}=R_t$. In other words,
\begin{equation}\label{eq:rhoSG} 
\rho_{\rm SG}= \frac{3}{32\pi}\frac{c^6}{G^3 M_{\bullet}^2}\ .
\end{equation}
This critical density is depicted in Figure \ref{fig:scales}. 

DM self-annihilations place an upper limit on the DM density of a clump by requiring that the clump does not self-annihilate within  a given time $t$. In other words: 
\begin{equation}\label{eq:rhocl} 
\rho_{\rm cl}\lsim \frac{m_\chi}{\langle\sigma v\rangle t} \ ,
\end{equation}
where $\langle\sigma v\rangle$ is the thermal velocity-averaged annihilation cross-section times the particle velocity, $m_\chi$ is the mass of the DM particle. 
Considering the DM distribution in galaxies, the relevant time-scale is typically the age of the system, which results in a saturated core density, $\rho_{\rm sat}$, at the center of a galaxy \citep{GS99,Lacroix+14,Lacroix+18}. 
Here, however, we adopt the dynamical timescale, $t_D$, of the self-gravitating clump \citep{Ali+16}, that describes a significant change to the clump due to its own gravity
\begin{equation}\label{eq:t_d} 
t_D=\sqrt{\frac{3}{4\pi G \rho_{\rm cl}}} \ .
\end{equation}
We note that the Hubble timescale is irrelevant here, because there is no need to require that the clump, or the binary, will survive for a Hubble time.
Thus, setting the time in Equation (\ref{eq:rhocl}) to be the above dynamical time, we can obtain an upper limit on the clump's density, for self-annihilating, self-gravitating clumps 
\begin{equation}\label{eq:rhoclmax} 
\rho_{\rm cl,max}\lsim \frac{4\pi G}{3}\left(\frac{m_\chi }{\langle\sigma v\rangle}\right)^2 \ .
\end{equation}
In Figure \ref{fig:scales}, we show this upper limit for $m_\chi=100$~Gev DM particles and adopting the canonical velocity-averaged annihilation cross-section times velocity $\langle\sigma v\rangle=3\times 10^{-26}$~cm$^3$~s$^{-1}$ \citep[][]{Jungman+96}.

From the comparison between $\rho_{\rm SG}$ and $\rho_{\rm cl,max}$, we find a minimum SMBH mass that can allow formation of a  self-gravitating spherical clump:
\begin{equation}\label{eq:Mlimit} 
M_{\bullet,{\rm lim}}\sim \frac{3 }{8\sqrt{2}\pi}\frac{c^3}{G^2} \frac{\langle\sigma v\rangle}{m_\chi } \ .
\end{equation}
For $m_\chi=100$~Gev DM particles, we find $M_{\bullet,{\rm lim}}\sim 4.3\times 10^7$~M$_\odot$. This is the mass at the crossing point between $\rho_{\rm SG}$ and $\rho_{\rm cl,max}$, depicted in Figure \ref{fig:scales}. Larger DM mass particles will result in smaller limiting masses.

\section{Clump Masses and Annihilation of DM }\label{sec:clump}

\begin{figure}
\centering
\includegraphics[width=\linewidth]{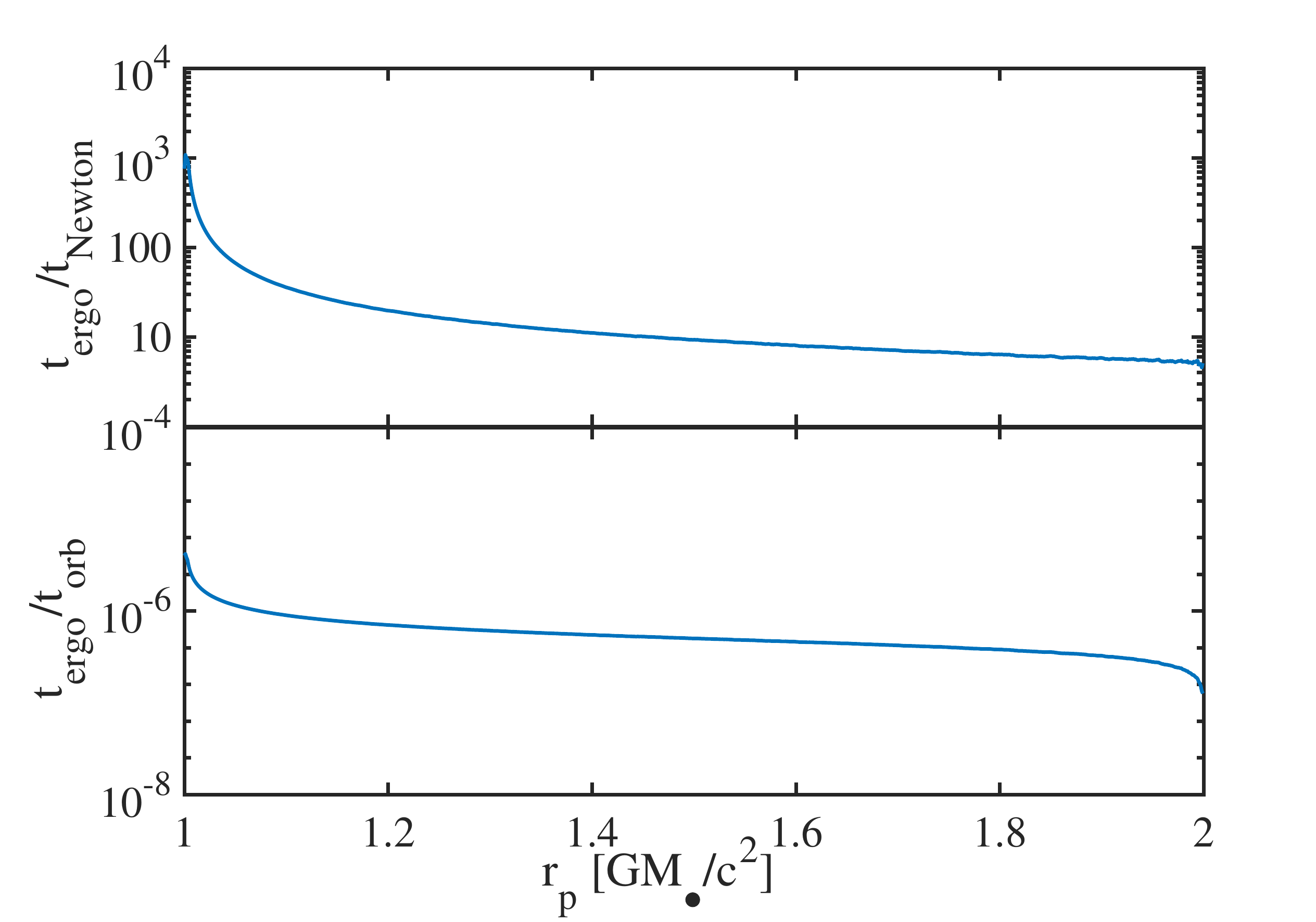}
\caption{ {\bf Top panel:}  we show the fractional orbital time a particle spends whirling in the ergosphere compared to the Newtonian time, $t_{\rm ergo}/t_{\rm Newton}$, as a function of the closest approach, $r_{\rm p}$, in terms of $G M_\bullet/c^2$.  {\bf Bottom panel} we show the relative time during the orbit the particle spends in the ergosphere, $t_{\rm ergo}/t_{\rm orb}$. In this estimate, the particles were placed initially at about $40,000 \times G M_\bullet/c^2$, which corresponds to the average location of the particles for a $10^7$~M$_\odot$ primary with a $10^9$~M$_\odot$ perturber from the Naoz \& Silk simulations. 
  }\label{fig:TimeErgo}
\end{figure}

Test particles, such as DM particles, on eccentric orbits that plunge  into the ergosphere, whirl around in the ergosphere a few times, then travel back out to apocenter [coined “zoom--whirl” orbits, because it involves several revolutions around the pericenter \citep{Glampedakis+02,Healy+09,Tsupko14}]. 
This behavior, of long-lived stable orbits, results in a density peak of DM inside $4G/c^2$ \citep[as shown numerically by][]{Schnittman15}. In Figure \ref{fig:TimeErgo}, we show the fraction of the orbit that a particle will spend zooming in the ergosphere compared to the Newtonian orbit (top) and as a function of the total orbit (bottom).    

To calculate the time spent in the ergosphere (Figure \ref{fig:TimeErgo}), we launch test particles on highly eccentric orbits with apocenter at $r=40,000M$, and pericenter ranging from $r=1M$ (the horizon) out to $r=2M$ (the outer surface of the ergosphere), where $M=G M_\bullet/c^2$. Along these Kerr geodesic trajectories, we simply integrate the coordinate time spent inside the ergosphere for each particle, and compare that to the total coordinate time of the orbit, and also the amount of time that would be spent inside of $r=2M$ with a Newtonian, non-relativistic orbit. From these, we see that the frame-dragging effects of the Kerr black hole lead to longer dwell times very close to the black hole, roughly a factor of $\sim 10-1000$ times longer than a Newtonian orbit. 

We note that in EKL systems, such as the one considered here, gravitational perturbations from the far-away companion excite the DM particles’ eccentricity to extreme values, for large range of inclinations and other orbital parameters \citep{Li+14,Li+13,NaozSilk14}. The eccentricity of these test particles continues to increase until they reach the horizon, and thus they pass through the entire ergosphere. Thus, Figure \ref{fig:TimeErgo}, depicts a typical behavior for particles that eventually will reach the ergosphere.

\citet{NaozSilk14} pointed out that  DM particles can reach extreme eccentricity, from grazing the ergosphere's outer surface all the way in to the inner surface of the ergosphere. These high eccentricity orbits result from  the gravitational perturbations from a companion that is  
far away from the primary SMBH, via the  aforementioned EKL mechanism.  The system is scalable and was shown to consistently result in nearly radial orbits \citep[e.g.,][]{Naoz+12bin,Li+13,Li+15}.  We note that even with systems for which the secular, double-averaged method\footnote{{The double-averaged method, averages over each of the orbits, and thus effectively the three-body system is reduced to two orbits,  interacting with each other  \citep[see for more details:][]{Naoz16}.}} breaks down, even larger eccentricities are expected \citep[e.g.,][]{Katz+12,Antognini+14,Luo+16,Hamers18}. Since the evolution is gradual, particles have their pericenters evolve from the outer to the inner surface of the ergosphere. In other words, at any given time  particles uniformly occupy the full extent of the ergosphere.

DM particles around an SMBH are on a stable orbit as long as $\epsilon=a_{\rm DM}e_{\rm p}/(a_{\rm p}(1-e_{\rm p}^2)) <0.1$  \citep[][]{Naoz16},
where  $a_{\rm DM}$ is the DM particle semi-major axis around $M_\bullet$, and $a_{\rm p}$ and $e_{\rm p}$, are the semi-major axis and the eccentricity of the perturber SMBH, respectively.
An eccentric SMBH perturber with  $M_{\rm p}>M_\bullet$ will excite the eccentricity of DM particles that are around $M_\bullet$ \citep{NaozSilk14}.  
 As a proof-of-concept, we set $a_{\rm p}$ to be roughly the sphere of influence, thus allowing for long-term surviving SMBH binaries. 
The shortest EKL timescale at which particles can reach $R_{\rm min}$ is proportional to  \citep{Naoz16} 
  \begin{eqnarray}\label{eq:tkozai}
\lefteqn{ t_{\rm EKL,min}\sim  \sqrt{\frac{a_{\rm p}^3}{M_{\rm p} G}} \sqrt{\frac{ M_\bullet}{M_{\rm p}}}\left( \frac{e_{\rm p}}{\epsilon}\right)^{3/2} \sim 10^3~{\rm yr} \times } \\ 
 &\times& \left(\frac{a_p}{1~{\rm pc}}\right)^{3/2} \left(\frac{10^9{\rm M}_\odot}{M_p} \right)\left(\frac{M_\bullet}{10^7{\rm M}_\odot}\right)^{1/2}\left(\frac{0.1}{\epsilon} \right)^{3/2}\left( \frac{e_{\rm p}}{0.8}\right)^{3/2} \nonumber \ . 
\end{eqnarray}
DM particles develop large eccentricities as time goes by until they reach such high eccentricities that they are consumed by the SMBH. 
We can thus scale Naoz \& Silk's nominal simulations, and estimate the time-scale at which ergosphere pericenter passages are expected. 
The time-scale 
has a similar time dependency for different SMBH mass primaries using the scaling relation in Eq.~(\ref{eq:tkozai}). 
We adopt this proof-of-concept time dependence for an assumed DM distribution within the SMBH sphere of influence. 

To estimate the mass that is temporarily accumulated on the ergosphere as a function of time we begin by calculating the DM density around the SMBH.
\citet{GS99} showed that the distribution of DM can be enhanced around the centers of galaxies, at a radius which is at the order of the SMBH sphere of influence. We adopt the DM density profile  inwards of the sphere of influence:
\begin{equation}\label{eq:densityprof}
\rho_{DM}=\left\{
\begin{array}{lc} 
0 & r\leq {2GM_\bullet}/{c^2} \ , \\
\rho_{\rm sat} & {2GM_\bullet}/{c^2} < r\leq R_{\rm sat} \ , \\
\rho_{\rm sat}\left(\frac{r}{R_{\rm sat}} \right)^{-\gamma} & R_{\rm sat}< r\leq R_{\rm spike} \ ,
 \end{array} \right. 
\end{equation}
where we assume a NFW profile for $r>R_{\rm spike}$, and $R_{\rm spike}$, {the radius at which DM density spikes}, is equal to the sphere of influence. The latter can be then computed for different SMBH masses via the $m-\sigma$ relation\footnote{{The $m-\sigma$ relation describes an observed correlation between the mass of the supermassive black hole at the centers of galaxies and the velocity dispersion $\sigma$ of the galaxy bulge.}} \citep{Tremaine+02}. From Eq.~(\ref{eq:rhocl}), the saturated density is $\rho_{\rm sat}={m_\chi}/({\langle\sigma v\rangle t_{\bullet}})$, where we adopt $t_{\bullet}=10^{10}$~yr as the age of the SMBH \citep{GS99,Lacroix+14,Lacroix+18}. The power-law index $\gamma$ is  expected to be between $2.25-2.5$ \citep{Boehm+09}, and in what follows we adopt $\gamma=7/3$ \footnote{We note that we have tested a larger value as well, which slightly changed the total clump mass but did not affect the overall, qualitative  conclusion. }. Demanding continuation between the different profile segments, we find the spike radius, $R_{\rm spike}$, and the saturated radius of the DM density, due to self-annihilation, $R_{\rm sat}$, (shown in bottom panel of Figure \ref{fig:scales}).

\begin{figure}
\centering
\includegraphics[width=1.1\linewidth]{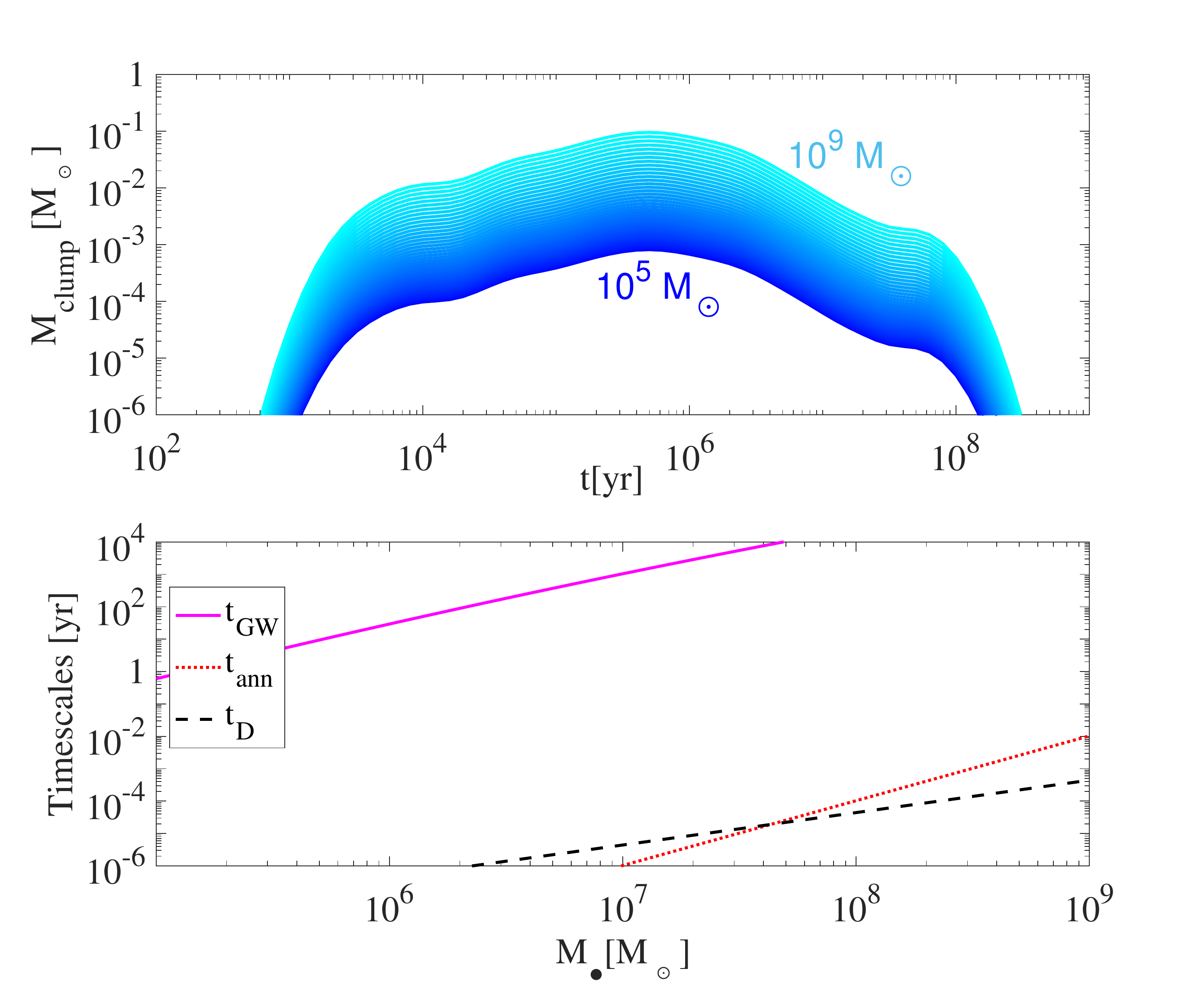}
\caption{ {\bf Top panel}: we consider the {mass of the self-gravitating clump as a function of time} at $R_{\rm min}$, for a range of {the primary} SMBH masses from  $10^9$~M$_\odot$  (cyan, top), to $10^5$~M$_\odot$, (blue, bottom). The time dependency is adopted from the Naoz \& Silk dynamical simulations, {where we assumed $e_{\rm p}=0.9$ and set $a_{\rm p}=R_{\rm Spike}$}. {The clump mass is estimated according to Eq.~(\ref{eq:Mclump})}. {\bf Bottom panel}:  the relevant time-scales in the problem. We show the GW time-scale, where we consider the maximum clump mass, and thus this is the shortest GW merger time-scale. We also consider the dynamical time-scale for a self-gravitating clump, Eq.~(\ref{eq:t_d}), (which is independent of annihilation processes). This time-scale represents a significant change that a spherical clump undergoes due to its own gravity.  Finally, we consider the annihilation time-scale (Eq.~(\ref{eq:tann})) which is much shorter than the GW merger timescale.   }\label{fig:Macc}
\end{figure}

 With the DM density profile at hand, we can now {\it roughly} estimate the mass of a self-gravitating clump. Using the results depicted in Figure \ref{fig:TimeErgo}, we find that  particles that reach the inner region of the ergosphere spend about $f_{\rm whirl}\sim 0.01\%$ of their orbital time near the ergosphere.  Thus, without loss of generality, we focus on these particles; as they develop their eccentricity over time, they swipe through the outer surface as well. Following \citet{NaozSilk14} we assume that about $f_{\rm acc}\sim30\%$ of all particles can reach the ergosphere\footnote{Note that this value can be as low as $15\%$ and as high as $50\%$. The former is a result of high eccentric particles that are captured by the massive SMBH perturber \citep{Li+15} and the latter is a result of having the perturber SMBH  grow in mass \citep{NaozSilk14}. }.  {Note that since these particles, by definition, reached the ergosphere and eventually the event horizon, they will reach the closest approach that corresponds to the $f_{\rm whirl}$ value we adopt from Figure \ref{fig:TimeErgo}.} Given different initial orbital configurations, different particles reach the required minimum pericenter at different times \citep[e.g.,][figure 9]{NaozSilk14}, since the eccentricity keeps on growing, the particles eventually are eaten by the SMBH. We scale this time dependency for different SMBH masses defining the fraction of particles with pericenter reaching the ergosphere inner surfaces as a function of time $f_{\rm EKL}(t)$ we can estimate the temporary clump mass as a function of time
\begin{equation}\label{eq:Mclump}
M_{\rm clump}\sim f_{\rm whirl}\times f_{\rm acc}\times  f_{\rm EKL}(t) \times M_{\rm available,DM} \ ,
\end{equation}
where $M_{\rm available,DM}$ is the available mass of the DM particles estimated within the stable regime using the DM density profile from Eq.~(\ref{eq:densityprof}). We show this mass as a function of time in the top panel of Figure \ref{fig:Macc}. The mass is decreasing after about $100$~Myr as particles either annihilate or are captured by the SMBH.

For annihilating DM particles, these high densities will undergo rapid annihilation. The  annihilation time-scale is estimated as
\begin{equation}\label{eq:tann}
t_{\rm ann}\sim \frac{m_\chi}{\rho_{\rm SG} \langle\sigma v\rangle} \ ,
\end{equation}
shown as dotted lines in the bottom panel of Figure \ref{fig:Macc}. We note that we use $\rho_{\rm SG}$ rather then $\rho_{\rm cl,max}$, because the former represents the minimum density at which the clump self-gravity will overcome the SMBH tidal forces, 
irrespectively of self-annihilation. However, the strong gravitational field of the SMBH may affect the dynamics, and thus Equation (\ref{eq:tann}) overestimates the annihilation time-scale for masses below $M_{\bullet,{\rm lim}}$. As the clump self-annihilates,  more particles can reach the ergosphere on  high eccentric orbits  via the EKL  mechanism  (as depicted on Figure \ref{fig:Macc}, top panel), thus forming a new clump, and the process rejuvenates.

\section{Gravitational Wave emission signal}

The orbit of a self-gravitating clump will shrink due to gravitational wave (GW) emission. We estimate the merger timescale  \citep{Peters64}, although we note that the clump is not a point mass, on the other-hand, due to the EKL, it does not exhibit radial symmetry either. We adopt the maximum clump mass estimated from Figure \ref{fig:Macc}, thus having a lower limit on the merger timescale between the clump and the SMBH (and the largest GW signal).  As depicted in  Figure \ref{fig:Macc}, bottom panel, the  merger time-scale via GW emission is much longer than the annihilation time-scale. Thus, while the clump undergoes self-annihilation, it emits a GW signal. In this example, the density around the SMBH will be replenished for about $10^8$~yr, yielding  continued GW and electromagnetic signals.

\begin{figure}[t!]
\centering
\vspace{-0.1in}
\includegraphics[width=\linewidth]{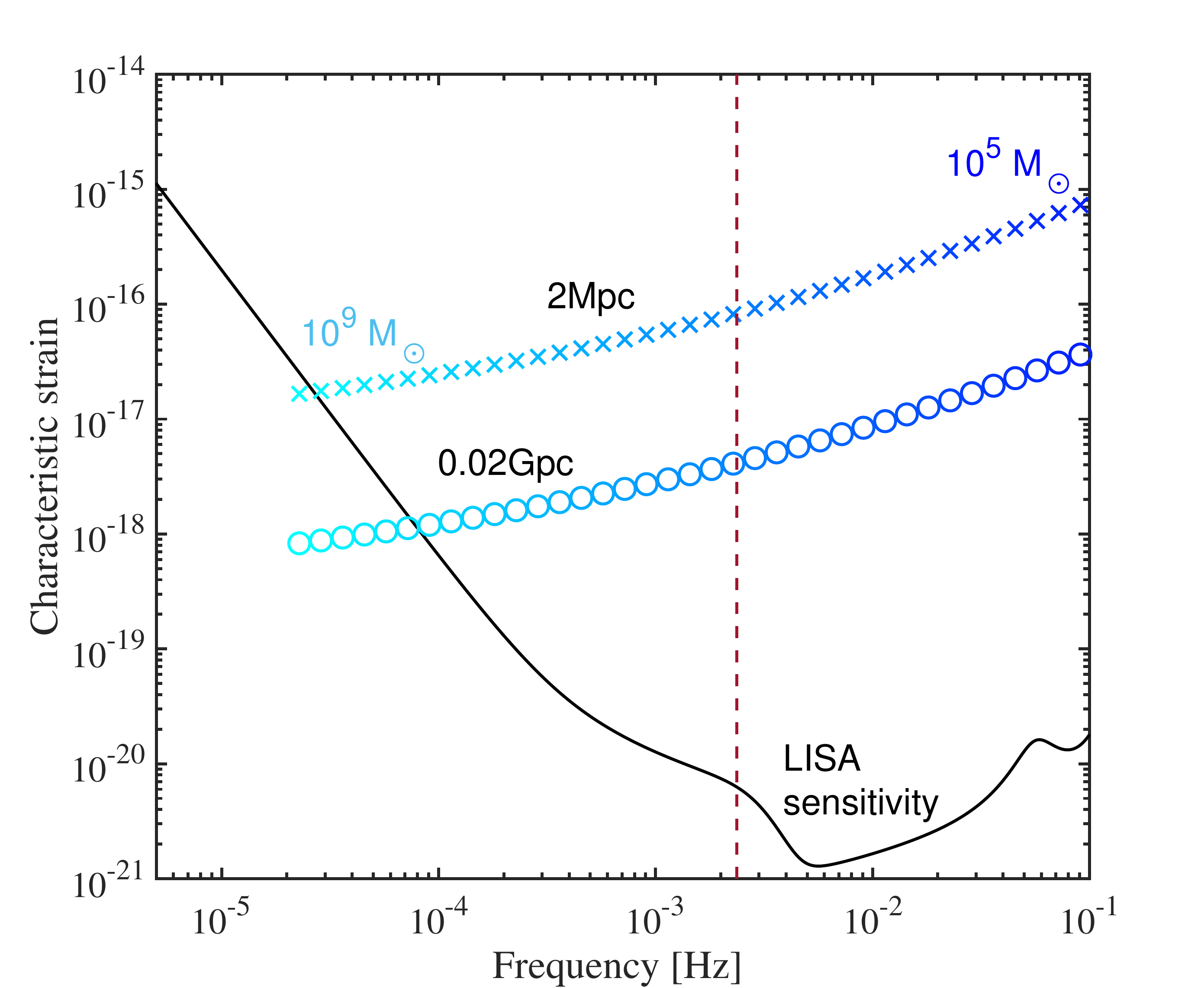}
\caption{ {\bf Examples for GW signal in LISA band.} We consider $4$~yr observational window  for source at $2$~Mpc and $0.02$~Gpc, top crosses and bottom circles, respectively. 
We show a range of SMBH masses from  $10^9$~M$_\odot$  (cyan, left), to $10^5$~M$_\odot$, (blue, right).  Over-plotted is the LISA sensitivity curve  \citep{Robson+18}. If DM self-annihilates, we find a limiting mass, $M_{\bullet,{\rm lim}}$, (corresponding to a maximum GW frequency) that can sustain the self-gravitating densities. We over-plot the corresponding GW frequency of this mass (dashed red line). {We remind the reader that the limiting clump mass, as well as the limiting GW frequency, depend on the DM mass and the annihilation cross-section, see Eq.~(\ref{eq:Mlimit}).}}\label{fig:LISA}
\end{figure}

We note that the spin axis of the SMBH orientation can have an arbitrary direction compared to the SMBH binary orbital plane. The DM particles clump in the ergosphere in an asymmetric configuration, and follow  highly eccentric geodesic trajectories through the ergosphere \citep{Schnittman15}. 
 Thus, the actual waveform of such a signal is rather complicated and depends on the orbital dynamics of a test particle in a strong gravitational field \citep{Will+17}. We therefore estimate the dimensionless characteristic strain for a circular orbit, which represents the order-of-magnitude of the expected signal \citep{Barausse+14}.  Moreover, most of the GW power is emitted near pericenter, so indeed the circular orbit approximation is a valid estimate \citep[e.g.,][]{Enoki+07,Tessmer+08}. 

The GW frequency $f$ of a circular orbit is twice the orbital frequency, and the dimensionless characteristic strain  is \citep{Robson+18}
 \begin{equation}\label{eq:Strain} 
h_c(a,f) = 2 h_0(a,f)f T_{\rm obs} \ ,
\end{equation}
where $T_{\rm obs}$ is the observation time window and $h_0(a,f)$ is defined as 
\begin{equation}
h_0(a) = \sqrt{\frac{32}{5}}\frac{G^2}{c^4}\frac{M_\bullet M_{\rm clump}}{D_l a} \ ,
\end{equation}
where $D_l$ is the luminosity distance and $a$ is the semi-major axis of the two objects. Below we adopt $a=R_{\rm min}$. Note that while these calculations do not include the spin of the SMBH, they are order-of-magnitude consistent with the latest implementation of the analytical model of Extreme Mass-Ratio Inspirals (EMRI) that does include the spin of a SMBH for point mass inspiral \citep{Chua+15,Chua+17}. We further emphasize that the strain calculation should be considered as strictly an order of magnitude, as orbits interior to the last stable orbit (such in our case) cannot orbit on circular orbits.

{Since the EKL mechanism yields an influx of DM particles on long timescales (e.g., Figure \ref{fig:Macc}), as a clump become self-gravitating and self-annihilates, additional DM particles may reach high eccentricites. Thus, we can consider a long observational window with LISA. } Figure \ref{fig:LISA} depicts the GW signal for two example sources observed for $T_{\rm obs}=4$~yr. The first is located $2$~Mpc away and the other located $0.02$~Gpc from us. We consider a range of SMBH masses (from $10^5$~M$_\odot$, to the right, to $10^9$~M$_\odot$ to the left). As can be seen in the Figure, LISA will be sensitive to a  large range of the SMBH mass parameter space. 

One may consider an observational window proportional to the annihilation time-scale. This time-scale covers a large range, from about a few days ($\sim 4$~d, for $10^9$~M$_\odot$) to less than a minute for the low mass SMBHs. This short time-scale is still within the LISA sensitivity window. It is unclear how LISA will handle very short observational windows.  However, as mentioned above, we expect a continuous formation of self-gravitating clumps of DM in the ergosphere, thus  short time-scales, which may be associated with burst-like signals, are less probable. 

\section{Discussion} 

We have shown that, thanks to the EKL mechanism in SMBH binaries, a self-gravitating DM clump may exist near the ergosphere of a spinning SMBH. DM may reach high eccentricities, spending considerable time there (see Figure \ref{fig:TimeErgo}), over long time-scales allowing for replenishing the possibly self-annihilating DM particles (see Figure \ref{fig:Macc}).  The mass of the clump can be high enough to allow for a GW signal, detectable by LISA (as depicted in Figure \ref{fig:LISA}). Our results suggest that this GW signal could be accompanied with a high energy electromagnetic signal from  DM self-annihilation processes,  and locked in phase to the GW signal, similar to the chirps predicted from neutron star-BH mergers  \citep{Schnittman+18}.

Note that if the DM does not self-annihilate into particles with detectable signatures, as is the case for gravitinos, we expect  continuous formation of a massive clump around the SMBH.
 Unlike the assumptions made in Figure \ref{fig:Macc}, where particles  either self-annihilate or are captured by the SMBH, if all the particles survive, the mass of the clump may increase over time.  Some of the DM particles may accrete onto the SMBH, however, the massive clump around the SMBH may result in, for example,   gravitino  decay products  \citep{Grefe12}. Moreover, the GW signal in such a case may be even stronger, as the clump mass may increase. However, in the case with no electromagnetic counterpart  we are unlikely to be able to distinguish between this GW merger and a stellar-mass extreme mass ratio inspiral. We note that we have assumed a gas-poor environment around SMBHs. Because the DM cusp dominates the density, and more so, as a clump becomes self-gravitating, we do not expect that interaction with the baryons will significantly alter the results.

SMBH binaries are a natural consequence of galaxy formation, consistent with observations \citep{Barrows+18}. We therefore expect that not only the torus-like DM distribution will be a generic outcome of SMBH binaries \citep{NaozSilk14}, but also have a GW signal from  self-gravitating DM that collects in the close vicinity of the SMBH. If DM self-annihilates, we predict that the GW signal will also be accompanied by a  high-energy signal.

\acknowledgements
SN and JS thank the UCLA Bhaumik Institute for Theoretical Physics for the hospitality that enabled the completion of this project. 
We thank the anonymous referees for useful comments that helped strengthen the paper.  
We also thank  Emanuele Berti, Bence Kocsis and Bao-Minh Hoang for useful discussions. SN acknowledges the partial support of NASA grant No. 80NSSC19K0321, and also thanks Howard and Astrid Preston for their generous support. JS acknowledges discussions with Enrico Barausse. JDS acknowledges support from NASA grant 17-TCAN17-0018.

\bibliographystyle{hapj}


\end{document}